\documentclass[11pt]{article}
\usepackage[margin=2cm]{geometry}
\usepackage{authblk}

\usepackage{caption}
\usepackage{subcaption}
\usepackage{graphicx}
\usepackage{siunitx}
\usepackage{lineno}
\usepackage{amsmath}
\usepackage{amssymb}
\usepackage{booktabs}

\newcommand{\dd}{\;\text{d}}

\usepackage{biblatex}
\bibliography{references.bib}

\begin{document}

\title{PoroTwin: A digital twin for a FluidFlower rig}
\author[1]{Eirik Keilegavlen\thanks{email:Eirik.Keilegavlen@uib.no}}

\author[2]{Eivind Fonn}

\author[2]{Kjetil Johannesen}

\author[3]{KristofferEikehaug}
\author[1]{Jakub Both}

\author[3,5]{Martin Fernø}
\author[2,4]{Trond Kvamsdal}
\author[2,4]{Adil Rasheed}

\author[1,5]{Jan M. Nordbotten}

\affil[1]{Center for Modeling of Coupled Subsurface Dynamics, Department of Mathematics, University of Bergen. 5020 Bergen, Norway.}
\affil[2]{Mathematics and Cybernetics, Sintef Digital, 7465 Trondheim, Norway}
\affil[3]{Department of Physics and Technology, University of Bergen. 5020 Bergen, Norway.}
\affil[4]{Department of Mathematics, Norwegian University of Science and Technology, 1344 Trondheim, Norway.}
\affil[5]{Norwegian Research Center, 5838 Bergen, Norway.}

\date{December 2022}
\maketitle

\abstract{
We present a framework for integrated experiments and simulations of tracer transport in heterogeneous porous media using digital twin technology.
The physical asset in our setup is a meter-scale FluidFlower rig.
The digital twin consists of a traditional physics-based forward simulation tool and a correction technique which compensates for mismatches between simulation results and observations.
The latter augments the range of the physics-based simulation and allows us to bridge the gap between simulation and experiments in a quantitative sense.
We describe the setup of the physical and digital twin, including data transfer protocols using cloud technology. 
The accuracy of  the digital twin is demonstrated on a case with artificially high diffusion that must be compensated by the correction approach, as well as by simulations in geologically complex media.
The digital twin is then applied to control tracer transport by manipulating fluid injection and production in the experimental rig, thereby enabling two-way coupling between the physical and digital twins.
}



\section{Introduction}\label{sec:introduction}
Insight into dynamic processes in porous media has traditionally been reached by direct observations and measurements of the processes combined with analysis and simulations of physics-based models (PBMs).
Continuous improvements in sensors and imaging give increased access to data of high quality and resolution.
In parallel, the field of data science, including data-driven modelling (DDM), artificial intelligence, and machine learning, has emerged as an alternative modeling approach that utilizes real-time data, and thus offers a complement to the physics-based models.
This calls for a hybrid approach, herein denoted Hybrid Analysis and Modeling (HAM), where data is integrated into physics-based in real-time using techniques from data science, enabling enhanced simulation accuracy and ultimately improved decision making in operational contexts. 
Such enhanced data models with tight integration between simulation models and the physical asset (the porous media), can be formalized trough the concept of digital twins \cite{Rasheed2020dtv}.
The purpose of this work is to present a digital twin which can access high-quality data from a porous media laboratory experiment, integrate this data into a physics-based simulation model in real time, and use the digital twin to guide decisions on the operation of the physical asset.

Porous media related data streams are undergoing a transformation with increased use of sensors and emerging imaging modalities. A range of imaging technologies provide data streams of high spatial and temporal resolution, which can be used to gain access to local fluid flows within the opaque porous media \cite{ferno2015combined,brattekaas2020unlocking}.
Recent focus on sub-Darcy-scale displacement processes utilize µCT \cite{berg2013real} and microfluidics \cite{gauteplass2015pore} to study pore-scale phenomena for a wide range of flow processes.
The increasing access to data is not limited to laboratory experiments:
Field-scale operation also increasingly rely on real-time monitoring of important processes, and regulations call for frequent seismic monitoring of carbon sequestration projects for prolonged time series \cite{arts2004seismic}.   
The amount of data generated can be substantial, with time-lapsed 3D or multimodal microscopic image sequences frequently generating several TB per experiment \cite{benali2022pore}. 
This calls for efficient approaches to incorporate data into analysis frameworks that can be used to harness physical insights contained therein.

One such framework is modeling and simulation based on physics-based models, which has a long tradition in porous media applications \cite{aziz1979petroleum,chen2006computational,helmig1997multiphase}.
The foundations of physics-based models, with a combination of fundamental physics, e.g., conservation or energy minimization principles, and constitutive laws, make them strong candidates for systematic studies and analysis of physical processes.
Data can be incorporated into simulation models in several ways, including long-loop parameter tuning where simulation models and setups are tweaked to honor observations \cite{williams1998stratigraphie}, and more continuous data assimilation methods which are popular in subsurface applications \cite{oliver2011recent,caers2011modeling}. 
Following trends in data science, data-driven models for porous media have recently been introduced, see for instance \cite{Ren2019Graph,Kierr2020flownet,Lie2022Graph}.
Compared to physics-based models, these data-driven models are computationally highly efficient, but not being based on physical principles, their use in interpreting physical phenomena is not straightforward.

\begin{figure}[!htb]
    \centering
    \includegraphics[width=\textwidth]{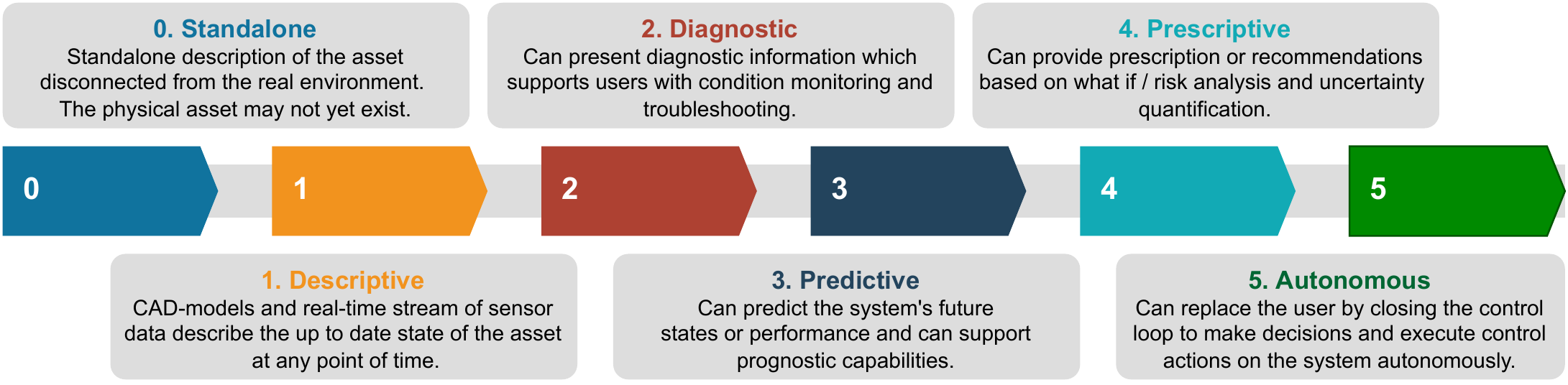}
    \caption{Increasing levels of capability for a digital twin. The capacity ranges from standalone, where the digital twin has no connection to the physical world, to fully autonomous, where the digital twin can control the physical twin without human intervention. From \cite{san2021hybrid}.}
    \label{fig:dt_capability}
\end{figure}

When used as a supplement, or substitute, for representing dynamics in a concrete physical asset (e.g., a porous media), simulation models can be interpreted as a digital twin of said asset. 
The digital-twin concept has been around for decades, overviews of applications to porous media can be found in \cite{wanasinghe2020digital,sircar2022digital}, while specific use cases include \cite{kannappin2022innovative,xiao2021permeability,ali2020virtual}.
The concept is very wide, in a sense any simulation model that aims to represent a specific porous media can be considered a digital twin. It is therefore relevant to classify digital twins in terms of their level of sophistication, and hence their ability to represent the status of, and inform decisions relating to, the physical asset.
As is illustrated by the classification shown in Figure \ref{fig:dt_capability}, this capacity can vary significantly. 

A critical component for enabling the higher capability levels is the digital twin’s ability to self-correct when provided with data that is incompatible with the current digital representation. 
In a sense, this is one goal of data assimilation and history matching techniques applied e.g. in subsurface engineering.
These techniques are however based on an assumption that all relevant physical processes are included in the governing equations and adequately represented in the discretized simulation model. If this is not the case, and it is often impossible to verify, the unresolved processes can only be represented by proxy.

Hybrid Analysis and Modeling offers an alternative version of self-correcting digital twins, which can account for mis- or underrepresented physics due to both poor parameter values and under-resolved physical processes.
The Corrective Source-Term Approach (COSTA) \cite{Blakseth2022dnn}, which falls under the HAM umbrella, employs an artificial neural network, trained on time series of combined observation and simulation data, to correct and enhance simulation results.
This enables us both to systematically include observations into a real-time simulation workflow and to use these observation to enhance the predictive accuracy of the simulations.

Herein, we describe the setup of a digital twin for fluid flow and tracer transport in porous media. 
The physical asset is a FluidFlower rig, which is essentially two-dimensional and  provides a data-rich environment through imaging and advanced image processing.
The digital twin consists of a COSTA-enhanced physics-based model, and the physical asset and digital twin are coupled using cloud technology.
Our contributions in this work can be summarized as follows:
\begin{itemize}
    \item We probe the ability of the digital twin to reproduce results from experiments without any feedback, thereby realizing a one-way coupling from the physical asset to the digital twin.
    \item We verify the ability of a data-driven model to correct unresolved and erroneous physics in a physics-based model. Moreover we present results for geologically complex media.
    \item We demonstrate two-way real-time communication between the physical asset and its digital twin, by considering real experiment in a meter-scale flow rig. In this case, we couple the digital twin to an optimization framework wherein well controls in the physical asset are manipulated to control the migration of the tracer plume, thus demonstrating the prescriptive capabilities of the digital twin.
\end{itemize}
As a result, the results presented herein demonstrate the feasibility of achieving level 4 autonomy (in the context of the classification scheme cited in Figure \ref{fig:dt_capability}). To the best of our knowledge no digital twin with similar capabilities has previously been presented for porous media.

The rest of the article is structured as follows: Section~\ref{sec:framework} explains the different components that build up the experimental setup and the digital twin. 
In Section \ref{sec:twin_accuracy} we present examples of one-way couplings from the physical to the digital twin, designed to probe the accuracy of the digital twin, while an example of a two-way coupling is shown in Section \ref{sec:application}.
Finally, in Section~\ref{sec:conclusion}, conclusions are given and potential future work is presented.

\section{Components of the PoroTwin framework}
\label{sec:framework}
\begin{figure}[!htb]
    \centering
    \includegraphics[width=\textwidth]{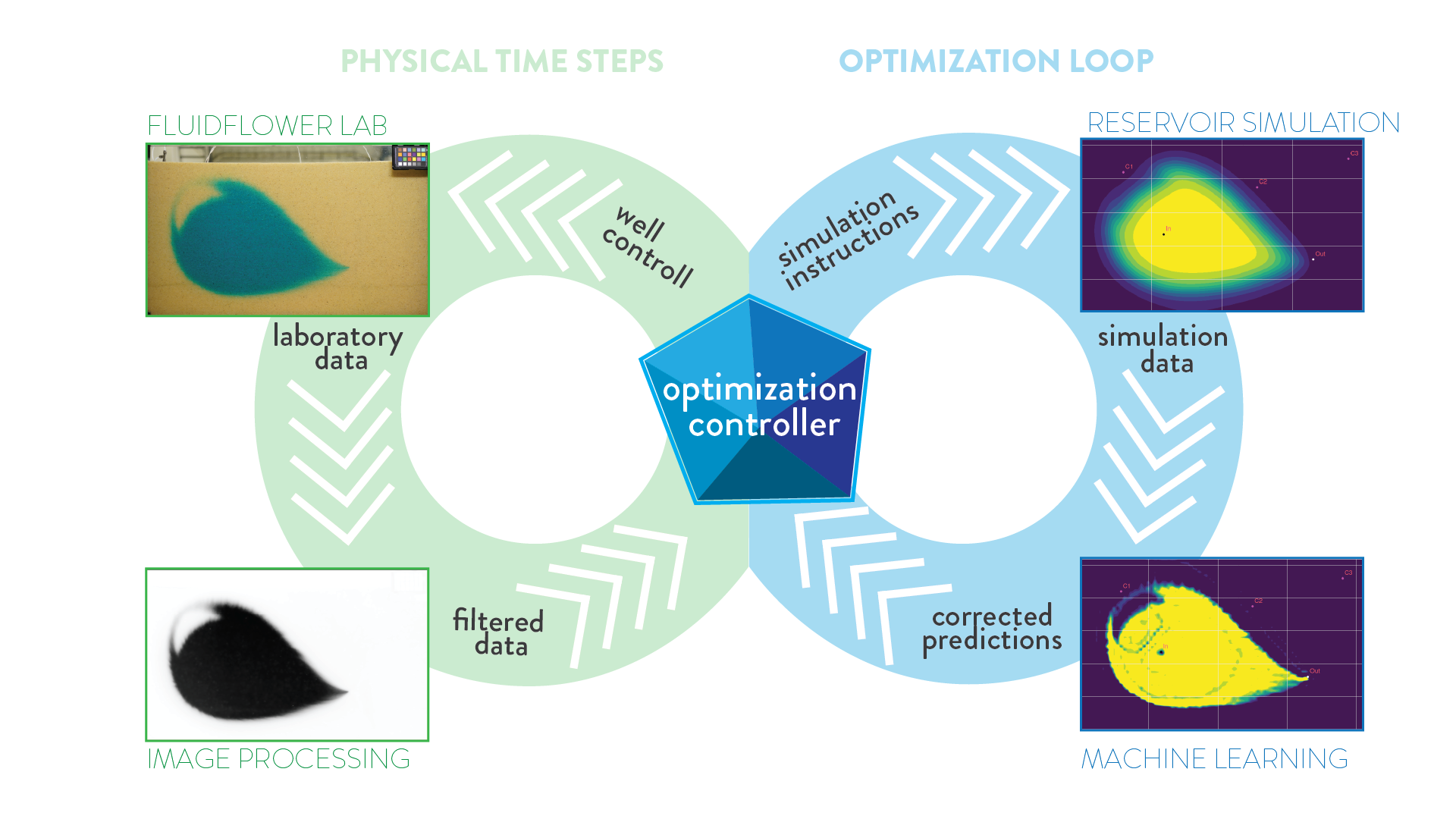}
    \caption{The components of the physical and digital twins, assembled for the optimization experiment presented in Section \ref{sec:application}. Data from the physical asset is filtered through image processing and sent to a 
    }
    \label{fig:optimization_workflow}
\end{figure}

Here we present the components that together make up the digital twin framework, termed PoroTwin: The physical asset, the digital twin, and the communication protocols that enable communication between the components.
The latter entails both flow of information, but also data post-processing, for example image processing.
To indicate how the components are coupled together, Figure \ref{fig:optimization_workflow} shows processes and information flows in the context of controlled injection of a tracer, as is studied in Section \ref{sec:application}.

The workflow as described below is to some extent tailored for the context of our physical asset and the experiments chosen in the results section, however the structure can be applied in a much broader context. Indeed, the well control and image data are considered laboratory scale proxies for well control and time-lapse seismic data typical for North Sea operations.  
Similarly, the digital twin can be employed for other purposes than optimization by modifying the central hub, while the other components remain relatively unchanged. This opens the door for considering risk assessment, 'what if?' analysis, uncertainty quantification, and process optimization. Furthermore, the digital twin can also be generalized to a set of digital siblings, i.e. a set of different physics-based simulators \cite{Rasheed2020dtv}.

\subsection{The physical asset}
\label{subsec:physicalasset}
Here we describe the physical asset in our setup, together with the data streams used in our experiments.

\subsubsection{The FluidFlower rig}
The FluidFlower enables meter-scale multiphase quasi two-dimensional flow experiments on model geological geometries with unprecedented data acquisition and repeatability.
The design allows for repeated injections tests with near identical initial conditions, allowing physical uncertainty and variability to be addressed using the same geological geometry.
This also enables well-confined sensitivity studies to evaluate large parameter spaces because each parameter may be varied while keeping others constant. 
The model geological geometry is constructed using unconsolidated sands and held in place between an optically transparent front panel and a technical back panel. 
The back panel has perforations that enable a range of well configurations (injector, producer, monitoring, or plugged) for porous media flow studies.
The optical access of the front panel facilities detailed investigation of flow processes are achieved through a high-resolution camera combined with a tracer. 

FluidFlower rigs of two different sizes are employed in this study:
A small rig with an extent of slightly less than a meter is considered in the experiments reported in Sections \ref{sec:costa_performance} and \ref{sec:application}, while a larger rig was used for the experiments reported in Section \ref{sec:geologically_complex}. 
For all experiments a passive tracer is used  to focus on the effect of well operation and, for the experiments reported in Section \ref{sec:geologically_complex}, how the flow is influenced by local geological features such as faults and multi-scale heterogeneity. 
Further details on the rig and the experimental setups are given in Section \ref{sec:twin_accuracy}.

Two types of data from the FluidFlower are made available to our digital twin: In addition to injection and production rates, the migration of the injected tracer is monitored using imaging as detailed next.

\subsubsection{Image processing}
\label{sec:image_processing}
The measurement instrument in the considered workflow is the combination of photographs taken of the physical asset and image processing. High-quality, high-resolution images provide dense data with low noise-to-signal ratio, as well as they resolve the medium up to sand grains. This high amount of details is beneficial in the further process. 
The high resolution on the other hand also provides a challenge, as it is important to emphasize that in the context of autonomous digital twins, all image processing must be fully automatic and real-time. 
As a whole, the role of the image analysis in this work is, in contrast to its typical applications, more a signal-to-data conversion, and less a tool for noise reduction. With that it is also the mediator of data from the physical to the digital twin.

\begin{figure}[t]
\centering
\makebox[\linewidth]{%
    \includegraphics[width=\textwidth]{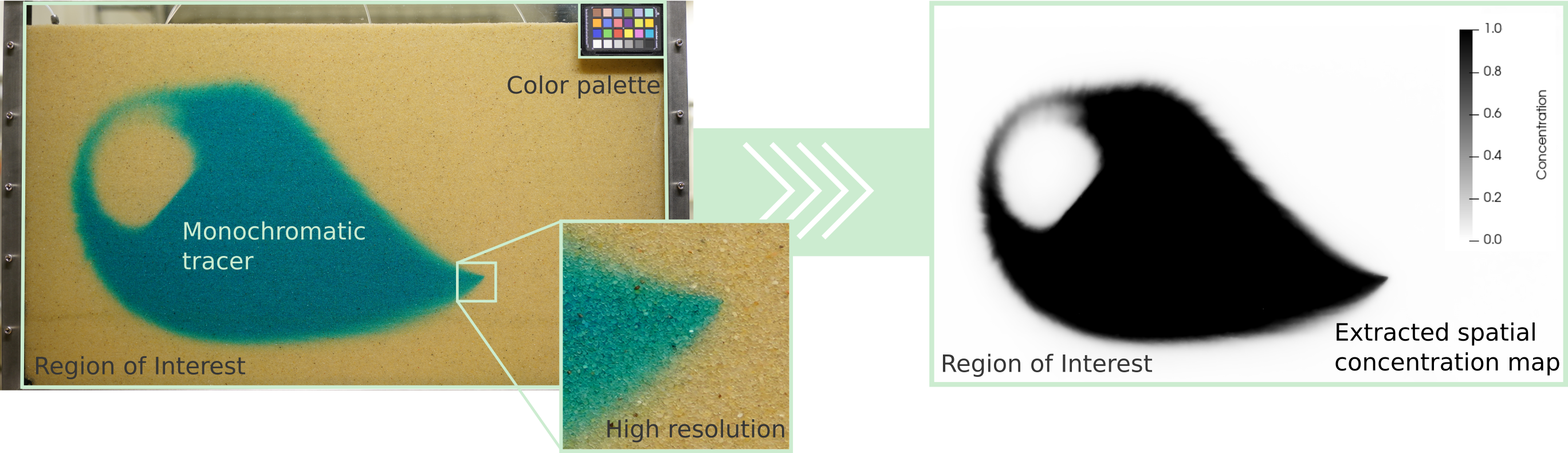}}
\caption{Illustration of the image analysis, from taken photograph to extracted spatial tracer concentration map.}
\label{fig:image-analysis}
\end{figure}

To convert a series of images of the conducted tracer experiments to spatio-temporal tracer concentration maps, the following steps are performed. First, a set of preprocessing routines is applied, including aligning all images with respect to a fixed point based on feature detection; cropping to a region of interest; and transforming the color spectrum such that the recorded colors of an attached color palette match a set of reference colors. Applying these to all images ensures a unified set of images. 
Next, to extract the tracer, the sand grains are removed by considering differences of images, using a baseline image taken before the start of the injection. Here the high quality of images comes into play. The removal of sand grains can be viewed as nearly noise free, yet, it leaves areas with little to no signal. We choose to apply total variation denoising~\cite{rudin1992nonlinear} as a combined tool for inpainting low-data regions (i.e. sand grains) and simultaneous denoising. Converting the images to a monochromatic color space provides scalar data in the interval $[0,1]$. The choice of the monochromatic color depends on the used tracer; here, grayscale images are considered suitable. 
Finally, a globally constant rescaling and clipping values at 1 is applied to convert the scalar signal to actual tracer concentration data. For this, a time series of images taken of a representative flow experiment with known injection rate is used for calibration. The final scaling parameter is obtained as the result of a RANSAC algorithm~\cite{fischler1981random}, aiming at matching the effective injection rate for the rescaled images, while excluding outliers. The result of the image processing is illustrated in Fig.~\ref{fig:image-analysis}.


Special care has to be applied in the context of heterogeneous media, as considered in Sec.~\ref{sec:geologically_complex}. As luminous emittance highly depends on the underlying material, the recorded signal becomes discontinuous over interfaces of different materials. Assuming piecewise homogeneous materials, materialwise constant scaling is applied, with values minimizing the signal jumps over interfaces in a least-squares fashion. Again, calibration is applied based on a representative set of images.

For the implementation of the image processing, \textit{DarSIA} (short for \textit{Darcy Scale Image Analysis toolbox})~\cite{darsia} is used, which is an open-source tool specifically developed for analyzing high-resolution images of porous media flow experiments, with integrated capabilities for converting data to continuum/Darcy scale. It includes both preprocessing routines as well as analysis tools for concentration- and deformation-based scenarios. 


\subsection{The digital twin}
\label{sec:digital_twin}

The digital twin in our setup combines traditional components of physics-based modeling and data-driven modeling in a hybrid analysis and modeling context as elaborated in the following section.

\subsubsection{Physics-based model}
The fluid flow in the porous media is modeled as incompressible and single-phase and is governed by 
\begin{eqnarray}
\label{eq:flow}
\mathbf{q} & = & -K\nabla p, \\
\nabla \cdot \mathbf{q} &=& f.
\end{eqnarray}
Here, $\mathbf{q}$ is the Darcy flux, $K$ is the hydraulic conductivity, $p$ is fluid potential and $f$ represents volumetric source and sink terms.
The transport of a passive tracer is given as
\begin{equation}
\label{eq:transport}
    \phi\frac{\partial c}{\partial t} + \nabla\cdot (\mathbf{q}c) = f_c,
\end{equation}
where $c$ is the tracer concentration, $t$ represents time and $f_c$ is tracer sinks and sources.

The above model ignores gravitational effects caused by minor density differences between water and the tracer, which may have an effect in the longer simulations presented in Section \ref{sec:geologically_complex}.
Moreover we model the FluidFlower as two-dimensional and thus ignore variations in the thin third dimension.

To solve Eqs. \eqref{eq:flow}-\eqref{eq:transport} in the forward simulator of the digital twin, we consider two different numerical tools: The toolbox IFEM is used for the experiments in homogeneous porous media presented in Sections \ref{sec:costa_performance} and \ref{sec:application}.
IFEM is an open-source object-oriented toolbox for implementing isogeometric finite element solvers for linear and nonlinear partial differential equations \cite{ifem_software}.
For the simulations in the geologically complex medium reported in Section \ref{sec:geologically_complex}, we employ finite volume methods as implemented in the open-source tool PorePy \cite{keilegavlen2021porepy}.

\subsubsection{Corrective source-term approach}
\label{sec:costa-motivation}
The COrrective Source-Term Approach (COSTA) is a methodology for combining the prediction from the physics-based simulator with data, in this case concentration maps from analyzed images, from the physical asset.
COSTA is designed to operate solely on source terms, while keeping the simulation model, intact.
In this section, we motivate the formulation of COSTA by considering a hypothetical setting where the true solution, in our setting the image data obtained through the analysis described in Section \ref{sec:image_processing}, is available.
In practice the purpose of the physics-based model, thus of the COSTA correction, is to predict an unknown future state based on information of the current state of the system.
To that end, a practical implementation of COSTA is described in Section \ref{sec:data-driven}.
More information on COSTA can be found in~\cite{Blakseth2022dnn}.

Let us represent the physics-based model as a (in this context linear) operator $\cal L$ over a domain $\Omega$  as 
\begin{equation}
    {\cal L}c:=\phi\frac{\partial c}{\partial t} + \nabla\cdot (\mathbf{q}c)
\end{equation}
Using the above operator Equation \ref{eq:transport} reads
\begin{align}
    {\cal L} \tilde{c} &=f_c, \quad \forall \ c \in \Omega 
    \label{eq:costaeq}
\end{align}
where $\tilde{c}$ represents the solution obtained by the physics-based model, which in practice will contain deviations from the true solution $c$ (in our case the image data).

It is instructive to consider potential sources of errors in the solution obtained from the physic-based model compared to a (hypothetical) solution operator for the continuous problem Equations \eqref{eq:flow}-\eqref{eq:transport}.
The solution obtained from physics-based model may carry the effect of incorrect parameters (e.g. the hydraulic conductivity $K$), poor modeling assumptions (e.g., neglecting three-dimensional effects), and numerical errors.
In addition, the physics-based model may be assigned an incorrect source term, in our case corresponding to mismatches between prescribed and actual injection and production rates. 
All these cases can be treated by adding a correction term $r$ to Equation \eqref{eq:costaeq}, and define the COSTA solution as
\begin{equation}
\label{eq:transport_costa}
    {\cal L}c_{COSTA}= f_c + r
\end{equation}
In practice the correction term will be identified by the DDM, however, in the hypothetical case where it is known exactly, we see that the COSTA solution is consistent with the true solution whenever the correction term corresponds to the residual error $r={\cal L}c - f_c$: 
\begin{equation}
\label{eq:costa_residual2}
    {\cal L}(c_{COSTA}-c)= f_c + r - {\cal L}c = (f_c-f_c) - ({\cal L}\tilde{c} - {\cal L}\tilde{c}) = 0
\end{equation}
For more information on COSTA and its treatment of different sources of errors, see \cite{Blakseth2022dnn}.

\subsubsection{COSTA implementation as a data-driven model}
\label{sec:data-driven}
Having motivated the COSTA correction term, we next turn to its practical implementation where the true solution $c$ is not known.
Specifically, to apply COSTA to the problem of tracer migration, the task is to compute $c_{COSTA}^{n+1}$ from a known state $c^{n}$ and a prediction $\tilde{c}^{n+1}$ obtained by the physics-based model.
Our aim is to use the digital twin to predict the future state of the physical asset in an application setting where $c^{n+1}$ is unavailable, thus the exact residual $r^{n+1} = \mathcal{L}( c^{n+1} -  \tilde{c}^{n+1})$ cannot be computed.

The COSTA approach relies on the availability of historical timeseries as training data to construct a data-driven model that approximates $r^{n+1}$ by
\begin{equation}
\label{eq:sigmadef}
    \sigma^{n+1} = DDM(c^n, \tilde{c}^{n+1})
\end{equation}
to the source term and apply these in \eqref{eq:costa_residual} so that this reads
\begin{equation}
\label{eq:costa_residual}
    {\cal L}c_{COSTA}^{n+1}= f_c + \sigma^{n+1}.
\end{equation}

The approximation $\sigma^{n+1}$ is implemented by training a deep convolutional neural network on training data generated from time series of tracer migration:
For each snapshot of the tracer, the physics-based model is invoked to predict the tracer distribution at the time of the next snapshot, and this is compared with the actual tracer distribution to compute a residual.
Using this training data, the data-driven model is constructed to correct for errors in the physics-based model, with the hope that the correction terms are accurate also for the tracer profiles encountered in the application phase.

\begin{figure}
\centering
\makebox[\linewidth]{%
    \includegraphics[width=\textwidth]{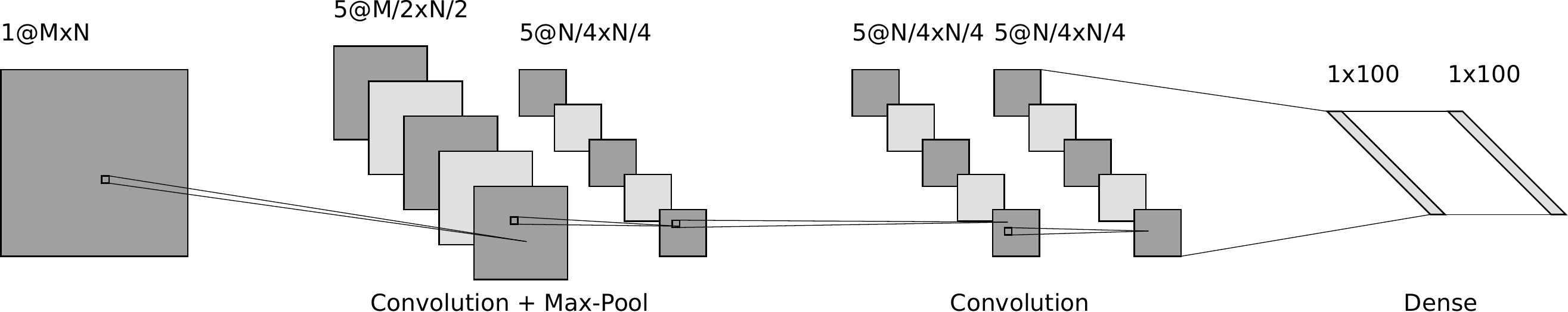}}
\caption{
Conceptual visualization of the neural network architecture used, see the text for a description of the individual components.
}
\label{fig:cnn}
\end{figure}

The convolutional neural network used in the current work is depicted in Figure \ref{fig:cnn} and can be expressed as:
\begin{equation}
    \mathbf{\hat{y}}=\mathbf{\hat{f}}(\mathbf{X})=(\mathbf{\Phi_Q} \circ \cdots \circ \mathbf{\Phi_2} \circ \mathbf{\Phi_1})(\mathbf{X}).
\end{equation}
In our case, the data $X$ is the predicted tracer profile $\tilde{c}^{n+1}$ and $\mathbf{\hat{y}}$ is $\sigma^{n+1}$.
The layers in the neural network, denoted $\mathbf{\Phi_i}$, may represent different operations, including convolution, activation and/or maxpooling.
As an illustration, in Figure \ref{fig:cnn} the input data represented by a matrix $\mathbf{X}$ of dimension $M\times N$ is subjected to $5$ convolution operations (filter dimension $5\times 5$) followed by $5$ max-pooling (filter dimension $2\times 2$) operations. After that follows a number of pure $5 \times 5$ convolution operations, without max-pooling. Finally, two dense layers consisting of $100$ nodes each are utilized. 
Within these dense layers, known parameters such as pump settings are concatenated.
The output layer (not shown in the figure) is of the same size and shape as the input. The precise number of layers used in each operation varies, and is clarified in the text as necessary. 

The activation function used in this work is ReLU. Since this is a regression problem, the mean-squared error (MSE) was utilised as a loss function to be minimized. The loss for the training batch $\cal B$ is then
\begin{equation}
    J_{MSE}({\cal B} ; \boldsymbol{\theta}) = \sum_{k\in{\cal B}} \Vert \mathbf{y}_k - \mathbf{\hat{y}}_k \Vert
\end{equation}
where $(\mathbf{x}_k, \mathbf{y}_k)$ is the $k$th pair in the dataset $\cal D$. For more detailed explanation of the CNN one is referred to \cite{Goodfellow2016dl}. The neural network was implemented using the tensorflow library \cite{tensorflow2015-whitepaper}.

\subsection{Information flow and communication protocols}
\label{sec:communication}
In addition to the physical and digital twin, our framework also constitutes a central hub which controls the action of the physical and digital assets.
Communication between the components is implemented using an \emph{internet-of-things} (IoT) model, with two types of agents and modes of communication:
\begin{itemize}
    \item \emph{devices} or \emph{servers}, which can emit one-to-many messages and respond to one-to-one queries from clients, and
    \item \emph{clients}, which can listen to one-to-many messages from devices, and make one-to-one requests.
\end{itemize}
The physical and digital assets are made available to the IoT network as devices. Other components of the system may optionally be published as IoT devices as well, such as the physics-based and data-driven models.
The central hub is implemented as a client. A typical purely predictive data flow may look like this:
\begin{enumerate}
    \item The physical asset emits a one-to-many message notifying listeners of a new system state.
    \item The hub receives it, and queries the physics-based model, using a one-to-one request, for a prediction from \eqref{eq:costaeq}.
    \item The hub in turn queries the data-driven model for a correction, using \eqref{eq:sigmadef}.
    \item The hub queries the physics-based model again for a corrected prediction from \eqref{eq:costa_residual}.
    \item The hub emits a suitably accurate future prediction to the digital asset.
\end{enumerate}
In a correction setting, the hub may instead make multiple queries for different control parameters (e.g.~well rates), and finally responding back to the physics-based model with updated and optimal settings. In Section~\ref{sec:application}, we present a case where the hub employs digital twin in an optimization framework to steer tracer migration in the physical asset, with a workflow illustrated in Figure~\ref{fig:optimization_workflow}.

The IoT model was realized using Microsoft Azure's IoT hub resources with cloud-based communication. 
Due to technical limitations with the pump-controlling software, the well controls in the physical twin could not be readily automatized, so the control messages are therefore communicated by audio.
The communication framework is implemented as open-source software and available at \cite{costa_software}.

\section{Accuracy of the digital twin}
\label{sec:twin_accuracy}
To establish the accuracy of the digital twin, including both the forward simulator, and its COSTA enhancement, we performed two sets of simulations.
The first probed COSTA's ability to compensate for unresolved physics, specifically to compensate for artificially high numerical diffusion.
The second considered simulations in a medium of complex geometry.

\begin{figure}
    \centering
    \includegraphics[width=0.48\textwidth]{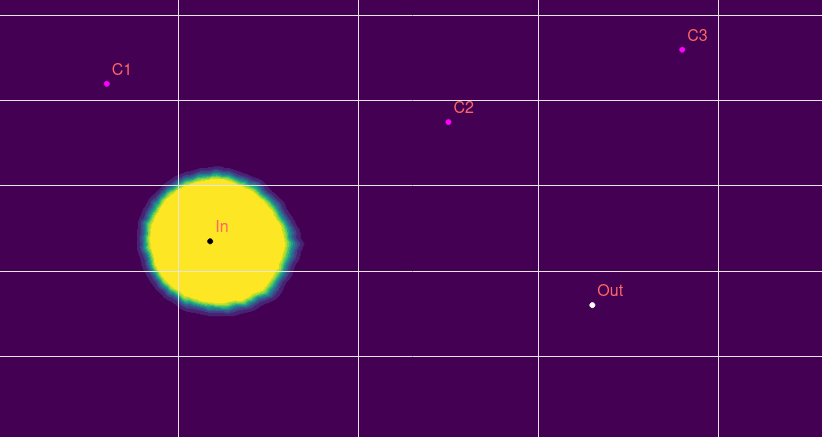}\\
    \includegraphics[width=0.48\textwidth]{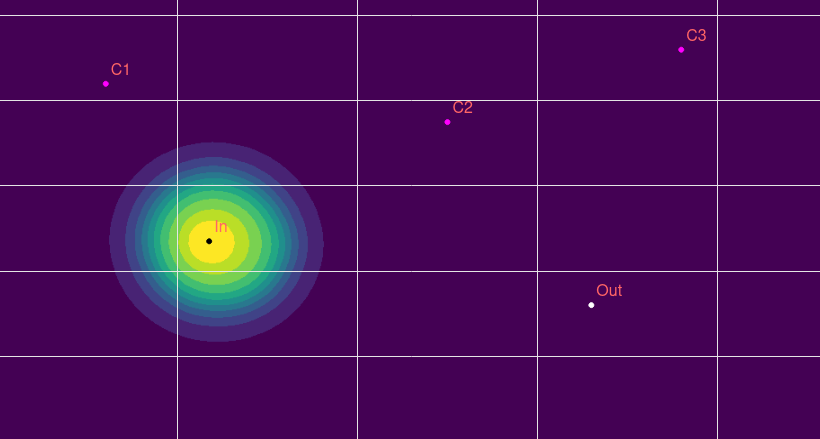}
    \includegraphics[width=0.48\textwidth]{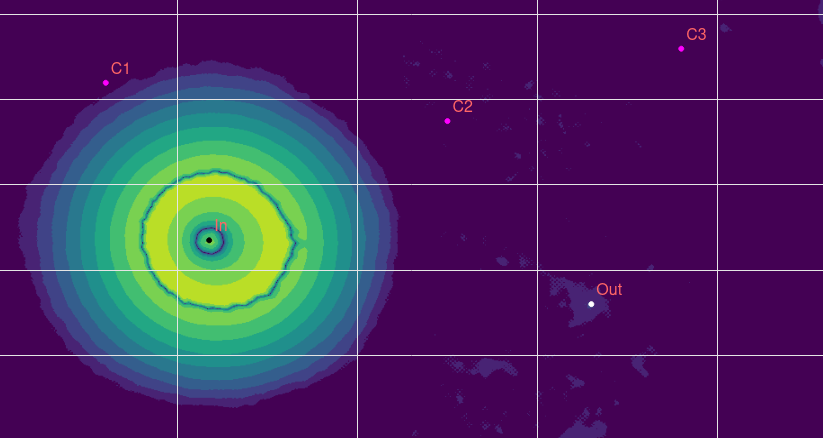}\\
    \includegraphics[width=0.48\textwidth]{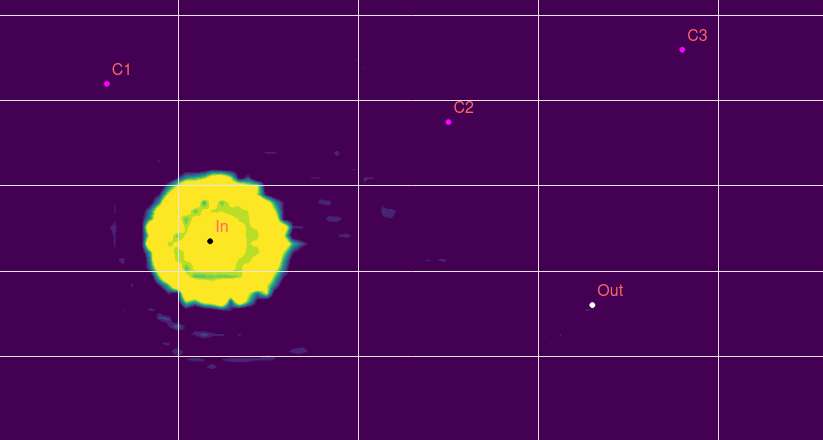}
    \includegraphics[width=0.48\textwidth]{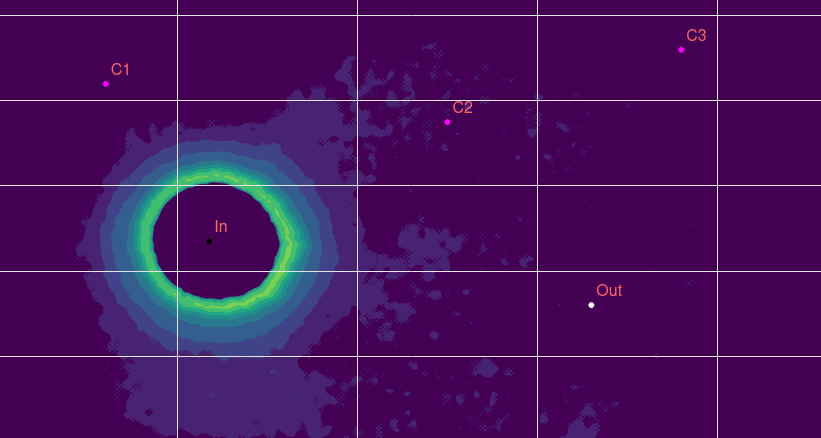}\\
    \caption{COSTA results at 10 minutes. At the top is the true experimental concentration
    field 30 seconds into the future. In the middle we see the predicted concentration field
    with pure IFEM (left) and the logarithmic error (right). At the bottom is the corrected
    COSTA prediction (left) and its logarithmic error (right).}
    \label{fig:costa-10}
\end{figure}

\begin{figure}
    \centering
    \includegraphics[width=0.48\textwidth]{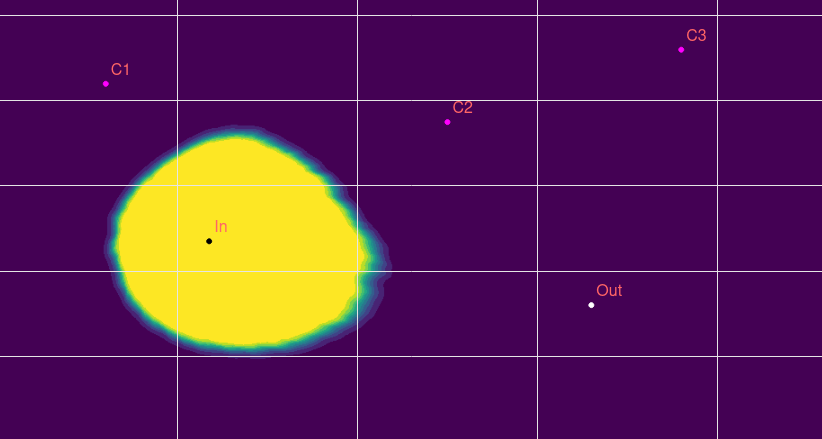}\\
    \includegraphics[width=0.48\textwidth]{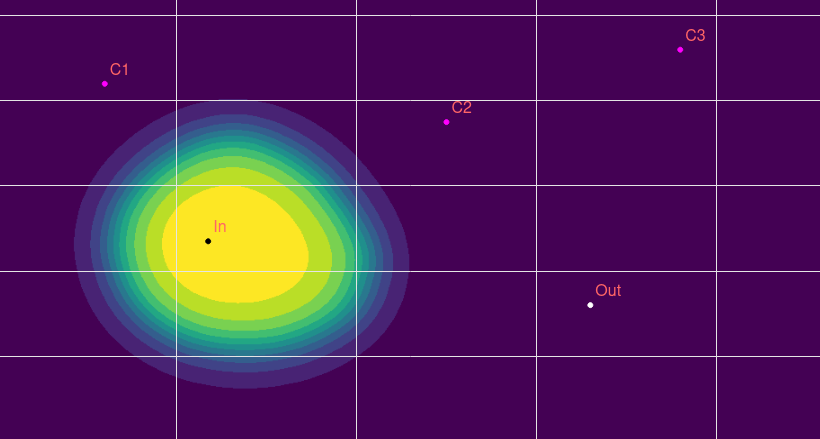}
    \includegraphics[width=0.48\textwidth]{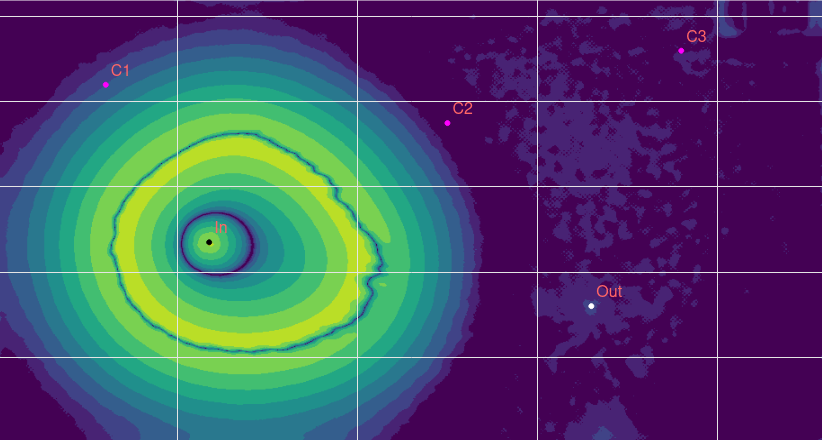}\\
    \includegraphics[width=0.48\textwidth]{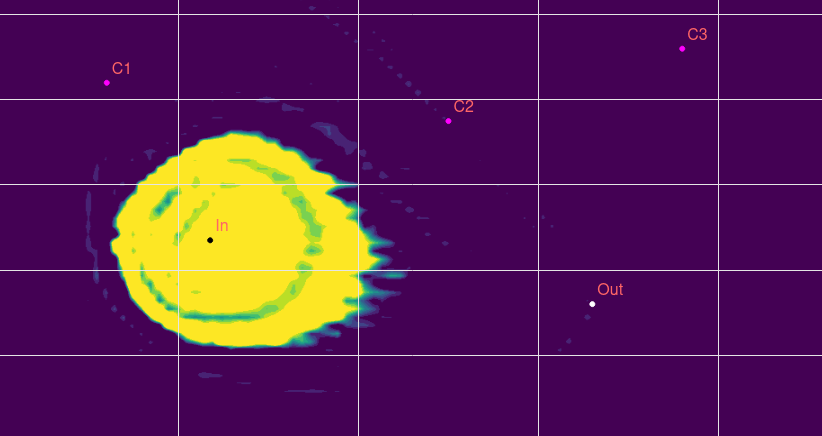}
    \includegraphics[width=0.48\textwidth]{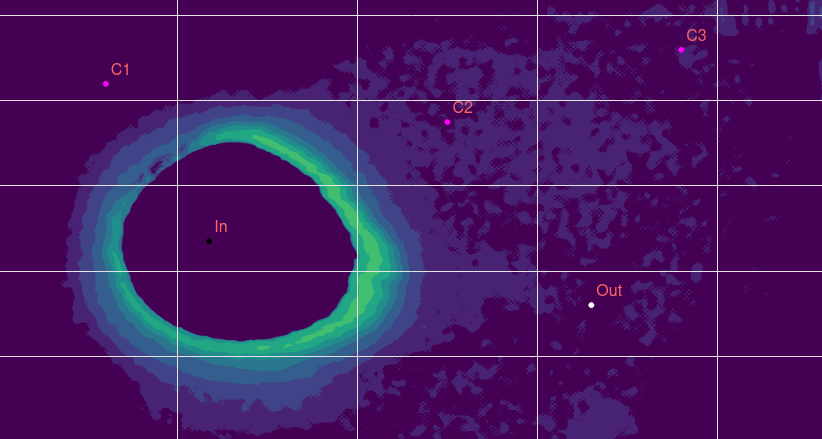}\\
    \caption{COSTA results at 30 minutes. See Figure~\ref{fig:costa-10} for explanation.}
    \label{fig:costa-30}
\end{figure}

\subsection{Performance of COSTA}
\label{sec:costa_performance}
We first considered simulations of tracer injection into a medium with homogeneous hydraulic conductivity. 
The numerical setup, using IFEM, was defined so that the physics-based model had unusually (and artificially) high numerical dispersion, and thus the predicted tracer profiles were overly diffusive.
A COSTA module was trained to compensate for this deficiency, with a goal of compensating for the poorly represented physics.
In addition to the known errors caused by high diffusion, there may have been other sources of errors as noted in Section \ref{sec:costa-motivation}, but these could not be measured nor distinguished from the diffusion-related errors.

The physical setup, used both for this experiment and in the experiment reported in Section \ref{sec:application} was as follows: 
The flow rig had dimensions $0.934 \,m \times 0.562 \,m$ in the $x$- and $z$-directions, respectively.
The extent in the $y$-direction was $0.010 \,m$, but for simulation purposes, the rig was considered a two-dimensional medium.
In total five wells were utilized in the flow experiments. One of these, termed \emph{producer}, 
was reserved for production. The others could be used for injection of tracer and water, they are called, respectively, \emph{injector} and \emph{control} 1 through 3, reflecting their role in the experiment reported in Section~\ref{sec:application}.
The locations of the wells are given in Table~\ref{tab:well_locations}.

\begin{table}[h]
	\centering
	\caption{Location of wells in the flow rig used for the experiments reported in 
    Sections~\ref{sec:costa_performance} and \ref{sec:application}. 
    The locations are given as offsets from the lower left corner of the rig.}
	\begin{tabular}{l|l|l}
		\toprule
		Well               & $x$ & $z$    \\
		\midrule
		Injector & 0.235 & 0.235 \\
		Producer & 0.660 & 0.160 \\
		Control 1 & 0.120 & 0.420 \\
		Control 2 & 0.500 & 0.375 \\
		Control 3 & 0.760 & 0.460 \\
		\bottomrule
	\end{tabular}
	\label{tab:well_locations}
\end{table}

For a given tracer distribution, the evolution the tracer plume depends on the rates of injection and production in the wells.
To train the artificial neural network underlying our implementation of COSTA, we collected tracer profiles from a range of experiments in the physical twin, injecting tracer in the different wells.
The resulting time series of tracer profiles, all taken 30 seconds apart, together with time series of the injection rates, were used to train COSTA following the protocol described in Sections \ref{sec:digital_twin}-\ref{sec:communication}. The neural network for this case was modeled according to the framework of Figure~\ref{fig:cnn}, with five combined convolution and max-pooling layers, no pure convolution layers and three pure linear layers.

The effect of the COSTA correction is illustrated for two different time steps, thus different tracer profiles, in Figures~\ref{fig:costa-10} and \ref{fig:costa-30}.
We observe that while the prediction quality for the tracer profile was poor due to to the artificially high diffusion, the corrected solution is remarkably similar to the actual state. 
Minor discrepancies can still be observed near the outer boundary of the predicted tracer profile, where COSTA failed to fully correct for the overly diffused profile, as well as in the region surrounding near the injection point, where the concentration reached a plateau of value 1.
Nevertheless, keeping in mind that the error in the predicted state was artificially large, the improvement in the simulation result from COSTA is notable.

\subsection{Performance in a geologically complex medium}
\label{sec:geologically_complex}
As a second test of the digital twin, we considered simulation of tracer injections into a domain with a complex hydraulic conductivity distribution.
Specifically we consider the geometry of the benchmarking exercise defined in \cite{nordbotten2022benchmark}, and aim to mimic the results 
of the well tests by tracer injection, which was part of the benchmark description.

\begin{figure}[ht!]
    \centering  
    \begin{subfigure}[t]{0.45\textwidth}
         \centering
         \includegraphics[width=\textwidth]{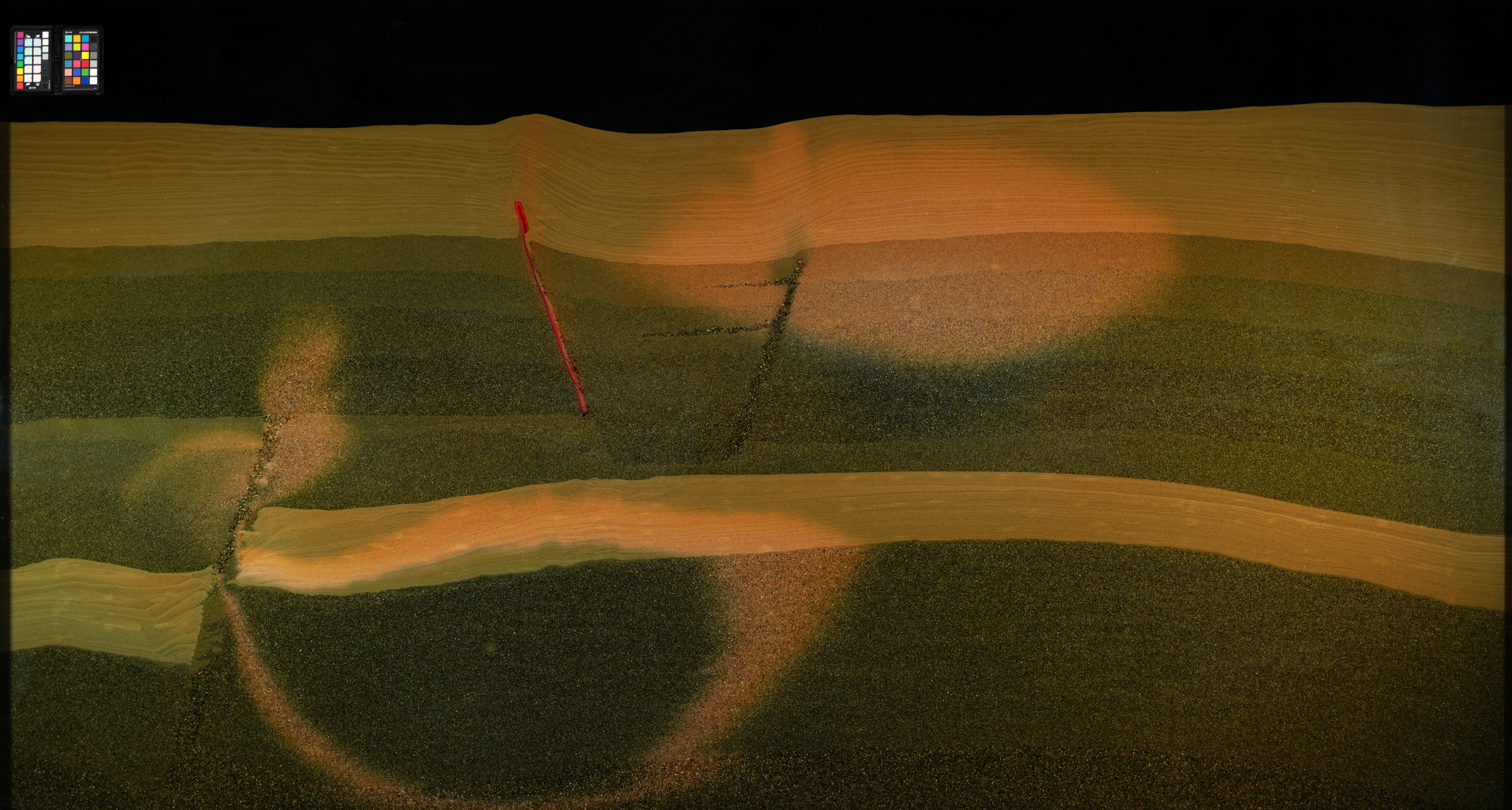}
         \caption{Final state of the tracer test.}
         \label{fig:complex-final-time}
     \end{subfigure}
     \begin{subfigure}[t]{0.45\textwidth}
         \centering 
         \includegraphics[height=3cm]{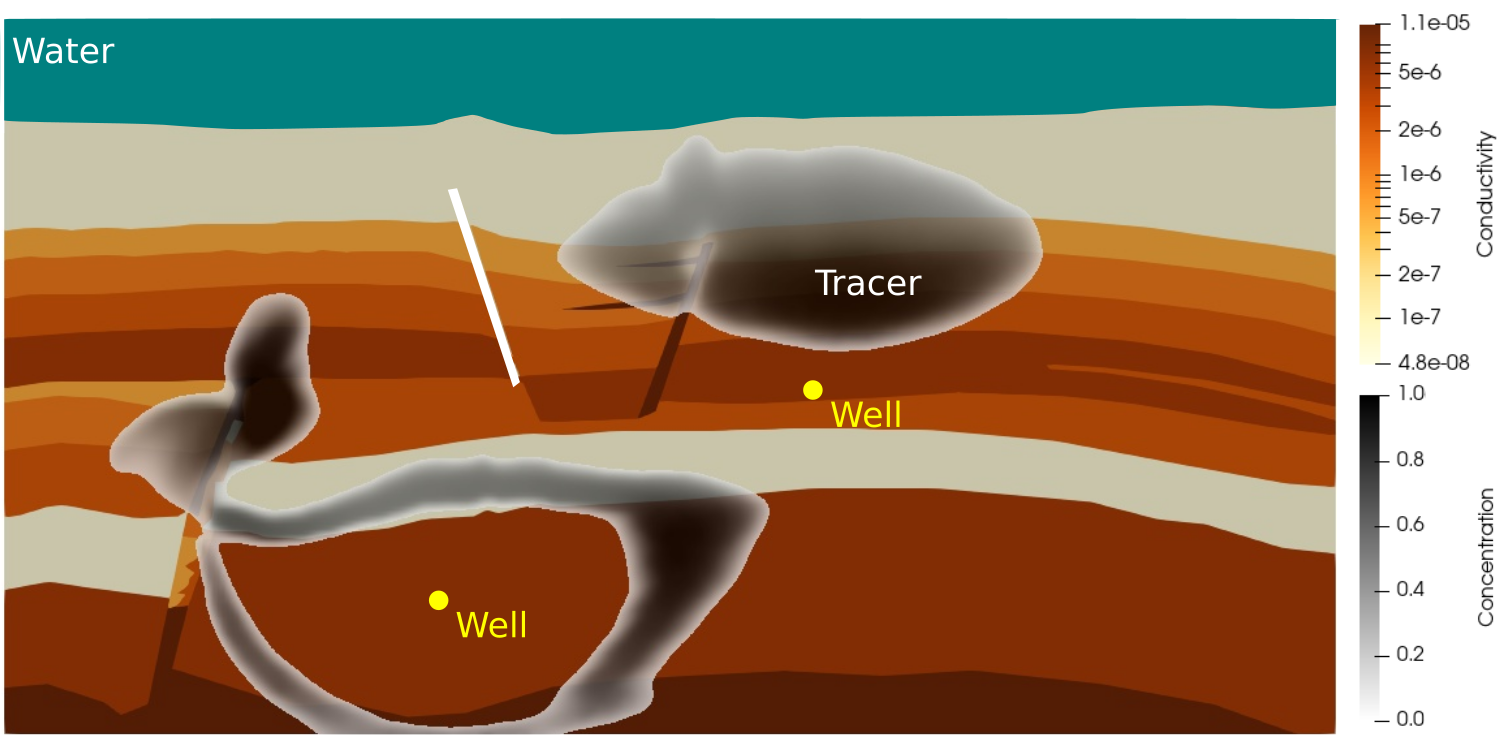}
         \caption{Concentration on top of conductivity.}
     \end{subfigure}
    \caption{Complex geometry and example photograph of the tracer test from the International FluidFlower benchmark study, together with the postprocessed concentration profile (cut off at small values for illustration purposes) on on top of the spatial conductivity distribution.}
    \label{fig:complex-image-analysis}
\end{figure}

In this case, the porous medium had a size of $\SI{2.8}{\meter} \times \SI{1.5}{\meter}$. 
The porous media were built from sands with different grain sizes, which translate into a hydraulic conductivity field that spans 2--3 orders of magnitude. The spatial distribution of the hydraulic conductivity field is visualized in Figure~\ref{fig:complex-image-analysis}, together with the location of the two injection wells used in the tracer test.
To account for the variable depth of this FluidFlower rig, see \cite{nordbotten2022benchmark}, the parameter fields were rescaled so that a two-dimensional simulation model could be applied.

\begin{figure}[h!]
\centering
\makebox[\linewidth]{%
    \includegraphics[width=\textwidth]{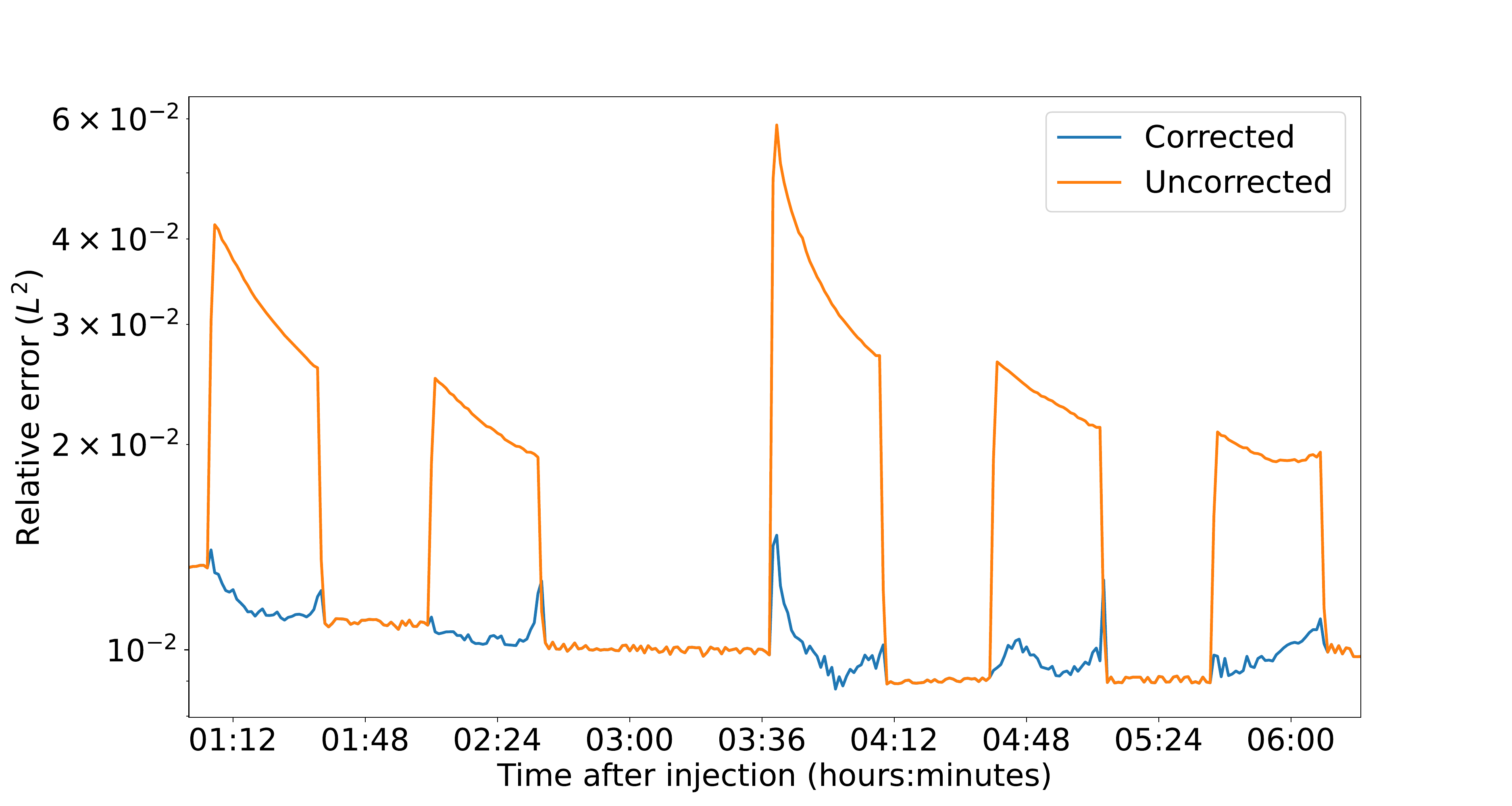}}
\caption{
    Uncorrected and corrected error in the geologically complex medium for \emph{one}
    timestep (i.e.~no accumulating time error). COSTA is able to correct the modeling errors
    made in certain time intervals.
}
\label{fig:complex-prederr}
\end{figure}

Tracer was injected in the two wells following the injection schedule described in \cite{nordbotten2022benchmark}. In short, with a total duration of 27.5 hours, tracer (partially alternating with water) has been injected in pulses of 30 minutes with longer breaks varying from 30 minutes to 14.5 hours; the final tracer distribution is displayed in Fig.~\ref{fig:complex-final-time}. Images were taken every 5 minutes providing the input data for the image analysis, which returns continuous concentration maps as described in Sec.~\ref{sec:image_processing}.
While the physics-based model assumes a perfectly passive tracer, in reality the tracer has a slightly larger density than water. During the longest injection break, notable displacement of the tracer can be observed in the physical experiment
Thus, buoyancy effects play a role which are not accounted for in the governing equations underlying the physics-based model.

A data-driven COSTA model was trained to correct single-timestep errors in this physics-based model. The model was built with
two downscaling convolution layers, five pure convolution layers followed by five dense linear `bottleneck' layers. In addition
to the pure image inputs from the physical asset, the data-driven model was also provided with the exact injection schedule as 
additional input to the dense layers. See Figure~\ref{fig:cnn} for details. 

\begin{figure}[ht!]
\centering
\makebox[\linewidth]{%
    \includegraphics[width=\textwidth]{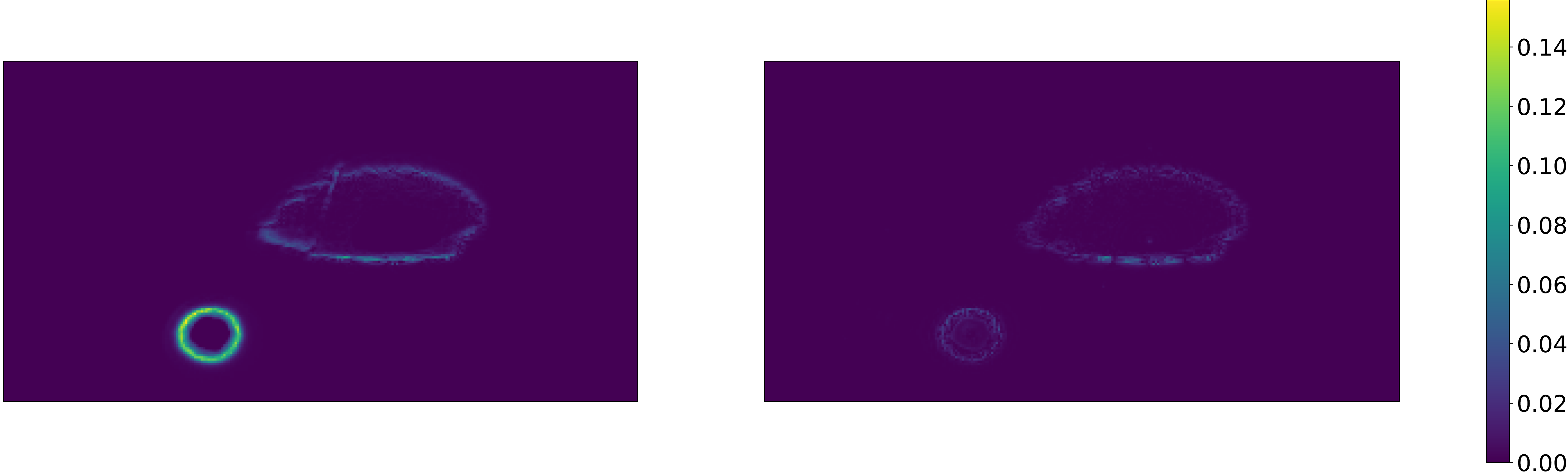}}
\makebox[\linewidth]{%
    \includegraphics[width=\textwidth]{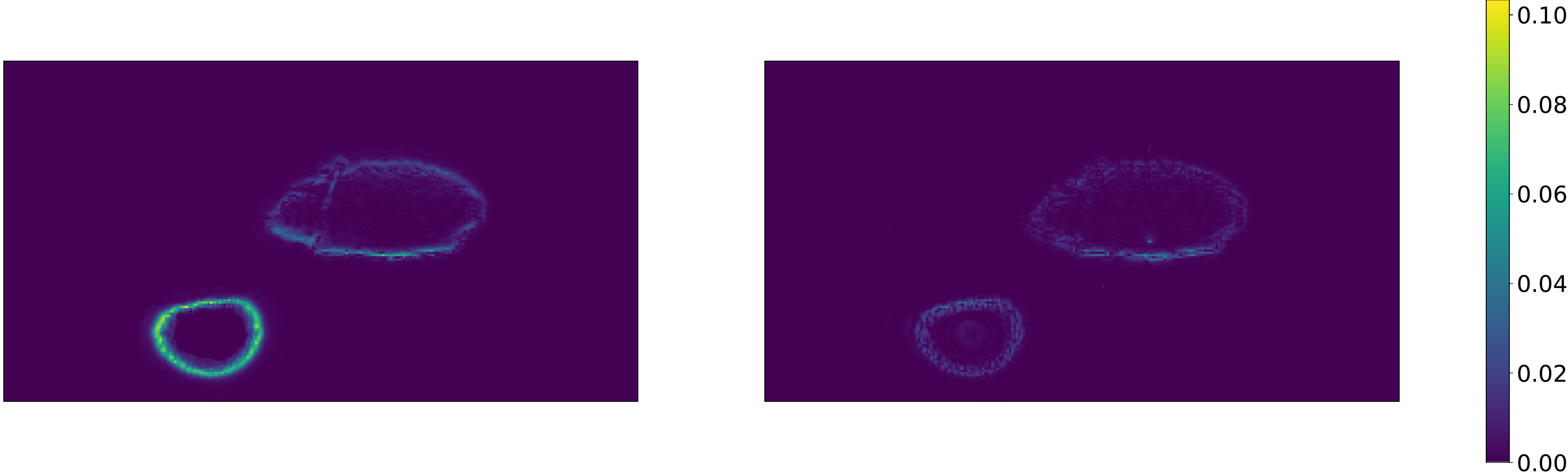}}
\caption{
    Errors for a single timestep, before correction (left) and after correction (right).
    Above: 3 hours and 50 minutes after start.
    Below: 4 hours and 40 minutes after start.
}
\label{fig:complex-spatial}
\end{figure}

The physics-based model is able to predict the next image with approximately $1\%$ relative error, with the exception of five distinct
time intervals within the one to six hour range after the start of injection, see Figure~\ref{fig:complex-prederr}.
Although some modeling error must be accountable for this effect, using COSTA we were able to correct the predictions in these intervals
to match the background noise error levels. See Figure~\ref{fig:complex-spatial} for spatial error distributions at two selected timesteps.

\begin{figure}[ht!]
\centering
\makebox[\linewidth]{%
    \includegraphics[width=\textwidth]{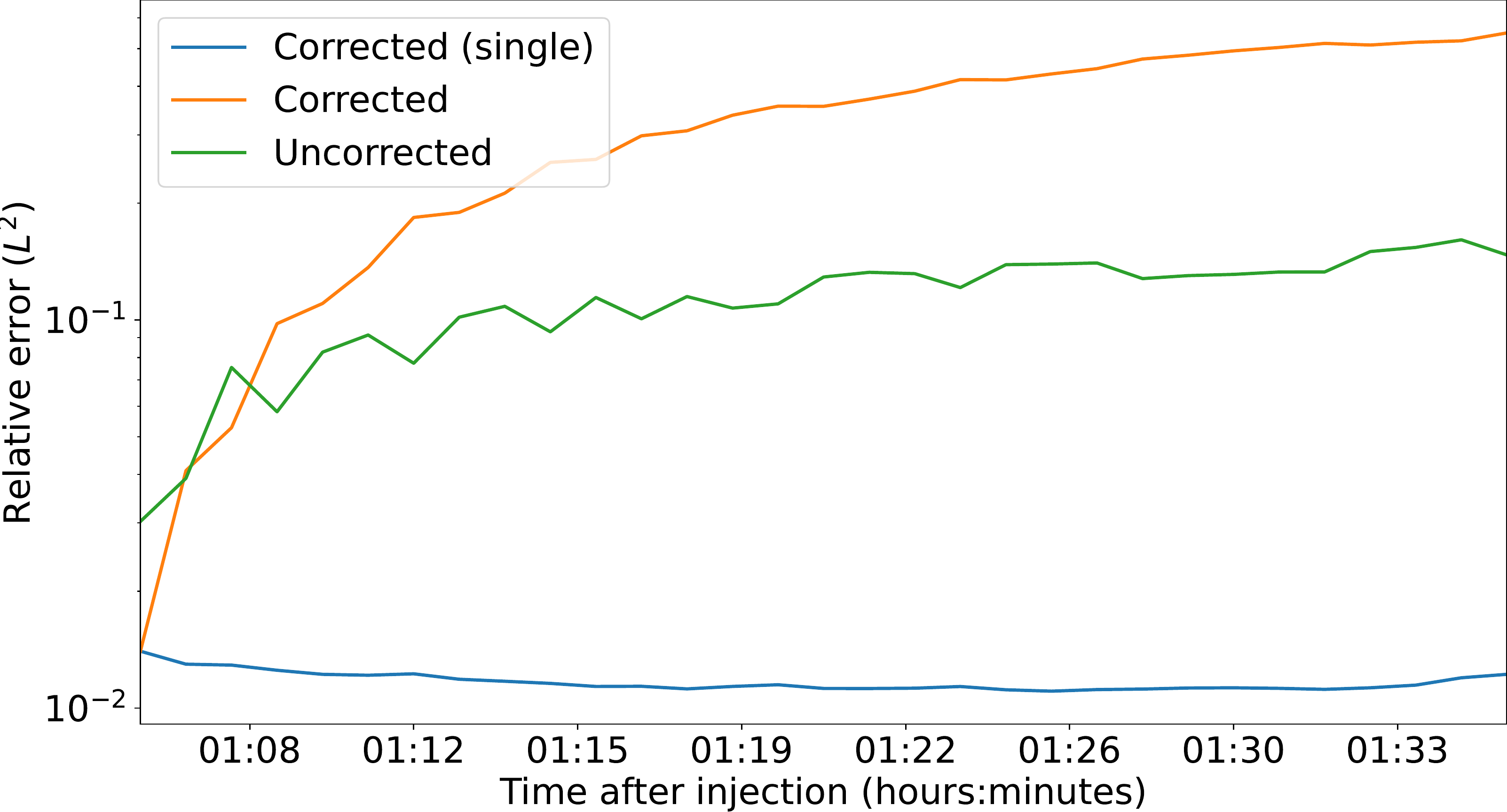}}
\caption{
    Accumulated time errors for the corrected and uncorrected timestepping,
    beginning at one hour and six minutes after start.
}
\label{fig:complex-accerr}
\end{figure}

Figure~\ref{fig:complex-accerr} shows the accumulated error accrued by long-term timestepping using both the uncorrected (PBM) and the 
corrected (COSTA) predictors, starting at one hour and six minutes into the experiment, coinciding with the first `jump' in PBM error.
This plot shows that, while the single-timestep corrected error remains low, COSTA faces challenges in generalizing the training set to perform corrections on initial conditions that are predicted and not sourced from `exact' image data. COSTA has shown itself capable of
doing this in other work, e.g.~\cite{Blakseth2022dnn}. Thus we attribute the deteriorating results in this case to potential overfitting on a narrow dataset.

\section{Application: Applying the digital twin for controlling tracer plume migration}
\label{sec:application}
The experiments presented in Section~\ref{sec:twin_accuracy} studied performance of the digital twin in a one-way coupling from the physical to the digital twin.
Here, we applied the digital twin to steer the physical twin, thus achieving a two-way coupling between the physical and digital twins, and arguably showing that the digital twin has reached capacity level 4 referring to Figure~\ref{fig:dt_capability}.
COSTA was set up using the same training data as described in Section \ref{sec:costa_performance}, but the injection pattern, thus tracer profiles, in this experiment was not present in the training set and not fed to the network.

\begin{figure}
\centering
\makebox[\linewidth]{%
    \includegraphics[width=0.47\textwidth]{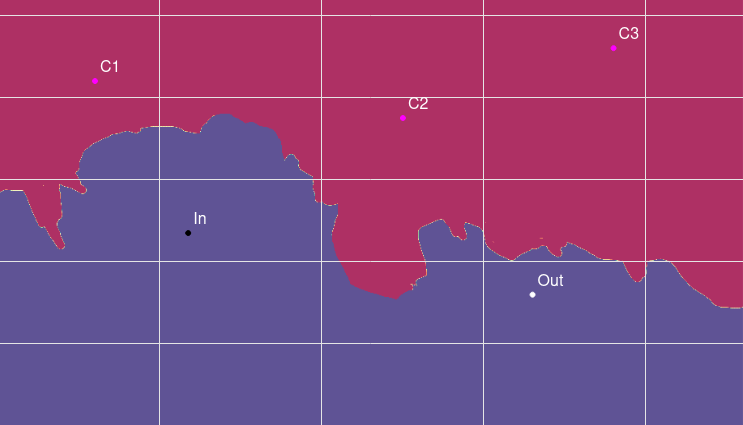}
    \includegraphics[width=0.47\textwidth]{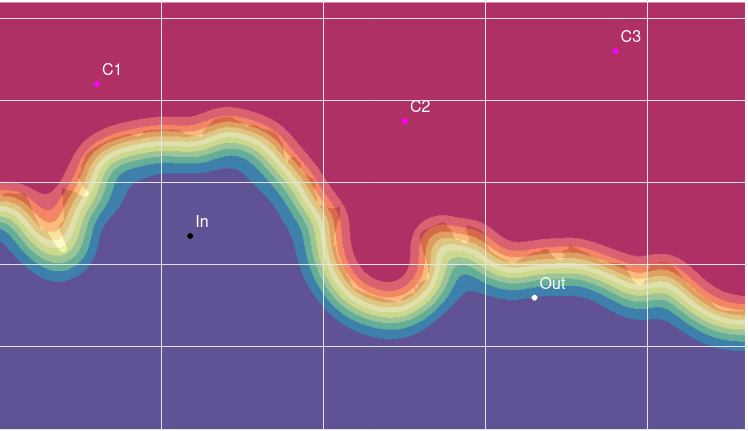}
    \includegraphics[width=0.03\textwidth]{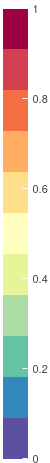}
    }
\caption{Disallowed zone for the tracer plume (left), and fuzzy mask $\varphi$ (right).}
\label{fig:nozone}
\end{figure}

We considered tracer injection into the same geometry as used in Section \ref{sec:costa_performance}, with the injection rate in well 1 fixed to $\SI{500}{\milli\liter\per\hour}$.
This resulted in a tracer plume, and the task of the digital twin was to steer the plume away from the upper part of the domain, with a demarcation line shown in Figure \ref{fig:nozone}. This corresponds to a hypothetical situation where injection of a waste (e.g. dissolved CO2) is controlled to stay within an operation licence. 

To this end, the digital twin was allowed to manipulate injection and production wells under the following constraints:
\begin{itemize}
    \item The injection rate for an individual control well could not surpass $\SI{500}{\milli\liter\per\hour}$.
    \item The \emph{total} injection rate could not surpass $\SI{500}{\milli\liter\per\hour}$.
    \item The production rate had to match the total injection rate so that the net injection rate was zero.
\end{itemize}
By corollary, the production rate was limited to $\SI{1000}{\milli\liter\per\hour}$, the sum of the fixed tracer injection rate and the maximal total control injection rate.

From the start of tracer injection, photographs of the tracer distribution were taken every $30$~seconds, and submitted to the digital twin. The digital twin then performed an optimization routine, minimizing a cost function
\begin{equation}
    f(c_1, c_2, c_3) = \int_\Omega c(x; \Delta t, c_1, c_2, c_3) \varphi(x) \dd x
\end{equation}
where $c_1, c_2, c_3$ are the injection rates for the three control wells subject to optimization, $c$ is the predicted future tracer distribution and $0 \leq \varphi \leq 1$ is a ``fuzzy'' map of the disallowed zone, see Figure \ref{fig:nozone}. This fuzzing was done with the intention of allowing the digital twin to respond with suitable control rates \emph{before} the tracer plume actually reached the disallowed zone, without necessarily needing to simulate so far into the future.

\begin{figure}
\centering
\makebox[\linewidth]{%
    \includegraphics[width=\textwidth]{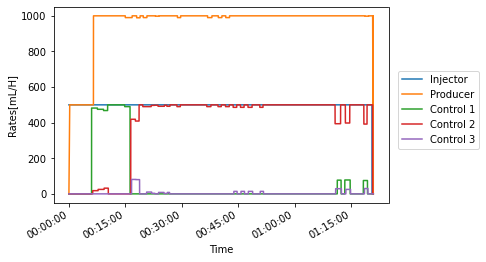}}
\caption{Well rates used in the optimization experiments. All rates refer to injection, except the producer. The injector rate is kept constant at $\SI{500}{\milli\liter\per\hour}$.}
\label{fig:well_rates}
\end{figure}

The minimization of $f$ was performed with the COBYLA method \cite{Powell1994ads}, a suitable algorithm able to handle all the constraints without requiring higher order data, such as derivatives. The implementation was provided by SciPy 1.5.4.

\begin{figure}
    \centering
    \includegraphics[width=0.48\textwidth]{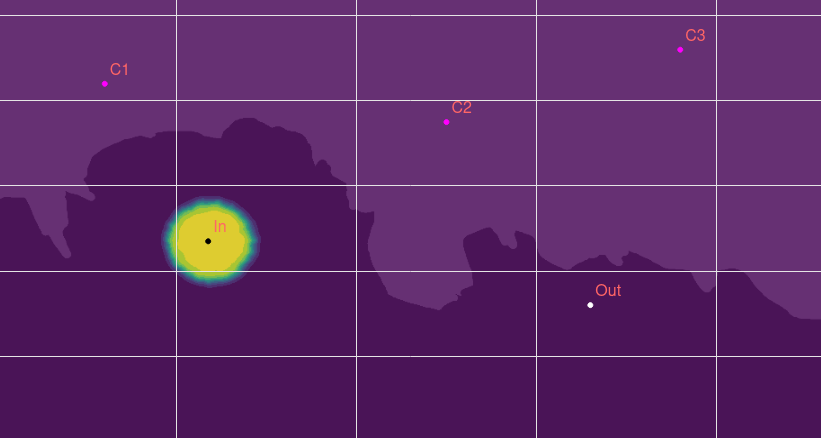}
    \includegraphics[width=0.48\textwidth]{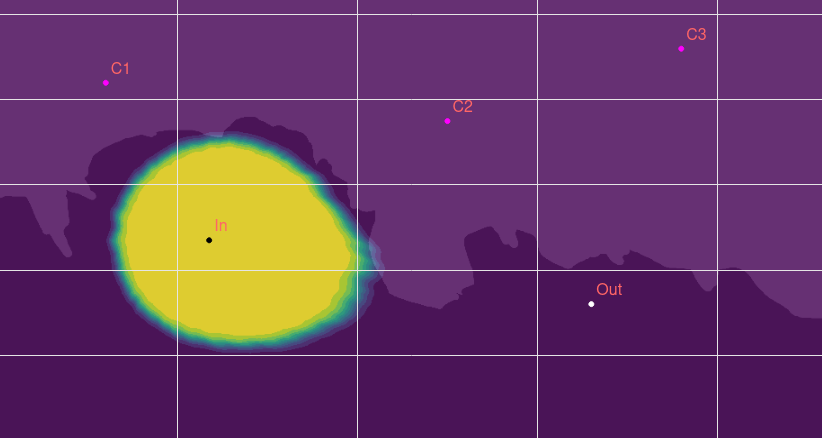}\\
    \includegraphics[width=0.48\textwidth]{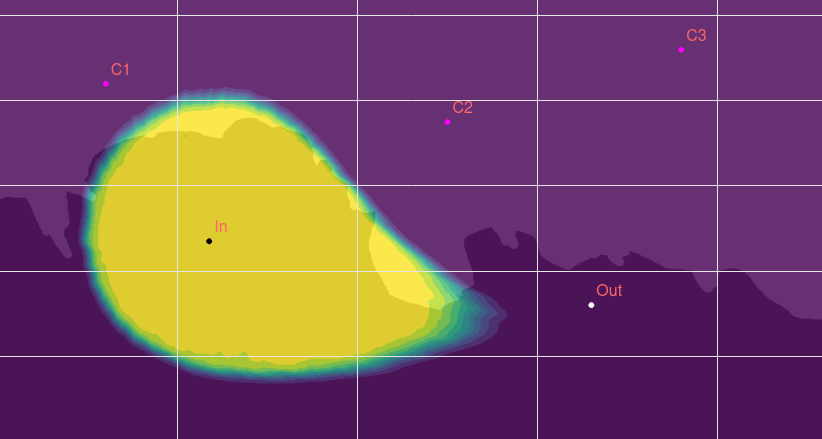}
    \includegraphics[width=0.48\textwidth]{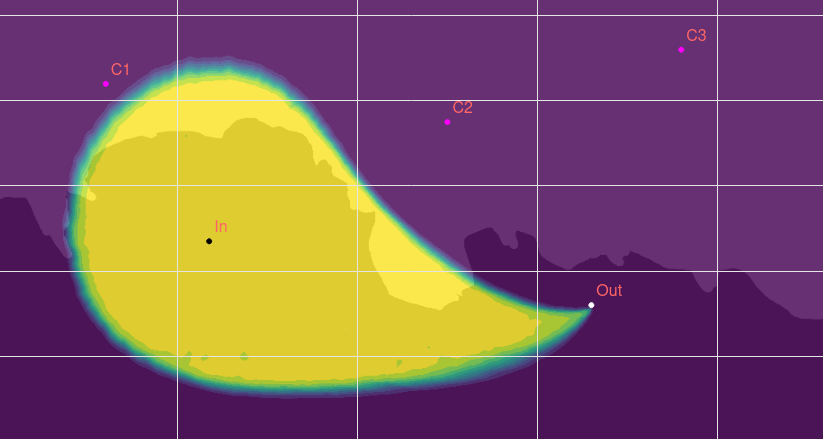}
    \caption{Evolution of the tracer plume, shown at, respectively, 10 minutes, 35 minutes, 60 minutes and 80 minutes from the start of the experiment.}
    \label{fig:plume}
\end{figure}

The implemented well control rates, as recommended by the optimization algorithm, are visualized in Figure \ref{fig:well_rates}, while the evolution of the plume is shown in Figure \ref{fig:plume}.
As can be seen, in the initial state the algorithm recommended controlling tracer migration through injection at close to the maximum allowed rate into Control 1, with some injection also into Control 2.
After about 15 minutes, the injection strategy switched to mainly prioritizing Control 2.
This pattern was mainly kept for the reminder of the experiment, interrupted by short periods of injection also into Control wells 1 and 3.
In practice, the measured production rate was at times approximately 1\% lower than the prescribed rate due to mechanical irregularities in the operation of the pump. This discrepancy was automatically accounted for in the digital twin through the source correction term described in Section \ref{sec:digital_twin}.

The impact of the control wells can clearly be seen in the evolution of the tracer plume:
Initially, injection in Control 1 leads to a deviation from the circular plume shape of an isolated injection.
The switch to injection in Control 2 seems consistent with an attempt to minimize migration into the disallowed zone, as does the reactivation of Control 1 late in the experiment.
Figure \ref{fig:plume} also shows the difficulty of the assigned optimization task, due to the combination of the geometry of the disallowed zone, the placement of the control wells, and the requirement that the tracer injection rate is fixed at 500mL/H.
Nevertheless, the example proves that our framework is capable of handling real-time interaction between the physical and digital twins.

\section{Conclusions and outlook}
\label{sec:conclusion}
We have presented a physical and digital twin for laboratory experiments of tracer transport in porous media.
The physical twin is a meter scale FluidFlower rig, while the digital twin consists of a traditional physics-based simulator augmented with a hybrid analysis and modeling (HAM) component.
The latter aims to capture physical effects that are not in the solution space of the simulator, and was in this case implemented using a corrective source-term approach (COSTA).
COSTA's ability to augment the physics-based simulator was illustrated with simulations in homogeneous and heterogeneous porous media.

We demonstrated the capabilities of the digital twin, and the possibilites in real-time communication between physical asset and digital twin, by controlling tracer migration in the physical twin by optimizing on well controls.
More broadly, we believe that the framework presented herein opens wide opertunities in combined experiments and simulations to study porous media dynamics in data-rich settings:
The tight integration offers a natural way to include data from observations into simulation models.
In contrast to traditional methods for data assembly, the COSTA framework need no assumptions on which parameters are uncertain, and it can enhance the range of the physics-based model to include effects not represented in the governing equations. 
This makes our approach an ideal tool for exploring and interpreting advanced laboratory experiments.

\section*{Acknowledgements}
This work was financed in part by Wintershall DEA through the project “A Digital Twin of Large-Scale Laboratory Flow Rig” (PoroTwin). 

\printbibliography


\end{document}